\documentstyle[ltwol,psfig]{article}

\def\be{\begin{equation}}
\def\ee{\end{equation}}
\def\bea{\begin{eqnarray}}
\def\eea{\end{eqnarray}}

\begin{document}

\title{MEASUREMENT AND PHENOMENOLOGY OF THE PROTON STRUCTURE FUNCTION
$F_2$ FROM ZEUS AT HERA}

\author{A. Quadt}

\address{Department of Physics, Particle Physics, 1 Keble 
Road, Oxford OX1 3RH, England\\E-mail: quadt@mail.desy.de}

%


\twocolumn[\maketitle\abstracts{ Measurements of the proton structure
function $F_2$ in the $Q^2$ range $0.6 - 17\;\rm GeV^2$ from ZEUS 1995 
shifted vertex data and $Q^2 \simeq 1.5 - 20000\;\rm GeV^2$ from 1996
and 1997 ZEUS data are presented. From the former and other ZEUS $F_2$ 
data the slopes $d F_2 / d \ln Q^2$ at fixed $x$ and $d F_2 / d \ln
(1/x)$ at fixed $Q^2$ are derived. $F_2$ data at $Q^2$ below $0.9\;\rm 
GeV^2$ are described successfully by a combination of generalised
vector meson dominance and Regge theory. 
Using a NLO QCD fit the gluon density in the proton is extracted in
the range $3 \times 10^{-5} < x < 0.7$ from ZEUS 1994 and 1995
data. For $Q^2
\sim 1\;\rm GeV^2$ it is found that the $q \bar{q}$ sea
distribution is still rising at small $x$ whereas the gluon
distribution id strongly suppressed. It is shown that these
observations may be understood from the behaviour of the $F_2$ and $d
F_2 / d \ln Q^2$ data themselves.  }]

\section{Introduction}
Measurements of the low and medium $Q^2$ \footnote{the negative of the
square of the four-momen\-tum transfer between the positron and the proton}
neutral current (NC) deep inelastic scattering (DIS) cross sections at
HERA have revealed the rapid rise of the proton
structure function $F_2$ as Bjorken-$x$ decreases below $10^{-2}$. 
At low $Q^2$ down to $0.1\;\rm GeV^2$ ZEUS data allows study of the
`transition region' as $Q^2 \rightarrow 0$ in which perturbative QCD
(pQCD) must break down.
At high $Q^2$, NC DIS measurements are sensitive to details of the QCD
evolution of parton densities, electroweak couplings and the
propagator mass of the $Z^0$ gauge boson. Furthermore, such
measurements allow the searches for physics beyond the Standard Model,
such as resonance searches or contact interactions.

\section{Phenomenology of $F_2$ at low $x$ and low $Q^2$}

\subsection{Phenomenology of the low $Q^2$ region}
The primary purpose is to use NLO DGLAP QCD on the one hand and the
simplest non-perturbative models on the other to explore the $Q^2$
transition region and through probing their limitations to shed light
on how the pQCD description of $F_2$ breaks down. One way to
understand the rise in $F_2$ at low $x$ is advocated by Gl\"uck, Reya
and Vogt (GRV94) who argue that the starting scale for the evolution
of the parton densities should be very low $(\sim 0.3\;\rm GeV^2)$ and
at the starting scale the parton density functions should be
non-singular. The observed rise in $F_2$, with a parameterisation
valid above $Q^2 \approx 1\;\rm GeV^2$, is then generated
dynamically. On the other hand, at low $x$ one might expect that the
standard NLO $Q^2$ evolution given by the DGLAP equations breaks down
because of the large $\ln (1/x)$ terms that are not included. Such
terms are taken into account by the BFKL formalism, which in leading
order predicts a rising $F_2$ at low $x$. The rise comes from a
singular gluon density, $x g \sim x^\lambda$, with $\lambda$ in the
range $-0.3$ to -0.5. Clearly accurate experimental results on
$F_2$ at low $x$ and the implied value of $\lambda$ are of great
interest.

At some low value of $Q^2$ pQCD will break down and non-perturbative
models must be used to describe the data. At low $x$ and large
$\gamma^*p$ centre-of-mass energy, $W\approx \sqrt{Q^2/x}$, the total
$\gamma^*p$ cross-section is given by
\begin{eqnarray}
\sigma_{tot}^{\gamma^*p}(W^2,Q^2) = \sigma_T + \sigma_L = \frac{4
\pi^2 \alpha}{Q^2} F_2(x, Q^2) 
\label{eqn:sigma_tot}
\end{eqnarray}
where $\sigma_T$ and $\sigma_L$ are the cross-sections for
transversely and longitudinally polarised virtual photons
respectively. Two non-perturbative approaches are considered, the
generalised vector meson dominance model (GVMD) and a Regge-type two
component Pomeron+Reggeon approach a la Donnachie and Landshoff (DL)
to give a good description of hadron-hadron and photoproduction total
cross-section data.

\subsection{Measurement of $F_2$ with Shifted Vertex Data}
The shifted vertex data correspond to an integrated luminosity of
$236\;\rm nb^{-1}$ taken in a special running period, in which the
nominal interaction point was offset in the proton beam direction by
$+70\;\rm cm$, away from the detecting calorimeter. Compared to the
earlier shifted vertex analysis, for the 1995 data taking period the
calorimeter modules above and below the beam were moved closer to the
beam, thus extending the shifted vertex $Q^2$ range down to $0.6\;\rm
GeV^2$.

The double differential cross-section for single virtual-boson
exchange in DIS is given by
\begin{eqnarray}
\frac{d^2 \sigma}{dx\, dQ^2} & = & \frac{2 \pi \alpha^2}{x\, Q^4}
\left[ Y_+ F_2 - y^2 F_L - Y_- x F_3 \right] \cdot
\left( 1 + \delta_r \right) \\
   & \simeq & \frac{2 \pi \alpha^2}{x\, Q^4}
\left[ 2(1-y) + \frac{y^2}{1 + R} \right] F_2 \cdot (1 + \delta_r), 
\label{eqn:doublediff}
\end{eqnarray}
where R is related to the longitudinal structure function $F_L$ by $R
= F_L / (F_2 - F_L)$ and $\delta_r$ gives the radiative corrections to
the Born cross-section, which in this kinematic region is at most
$10\%$. The parity violating term $x F_3$ arising from the $Z^0$
exchange is negligible in the $Q^2$ range of this analysis.  Further
details about the data analysis can be found in ref. \cite{low_pheno}.

\begin{figure}
\center{
\hfill 
\psfig{figure=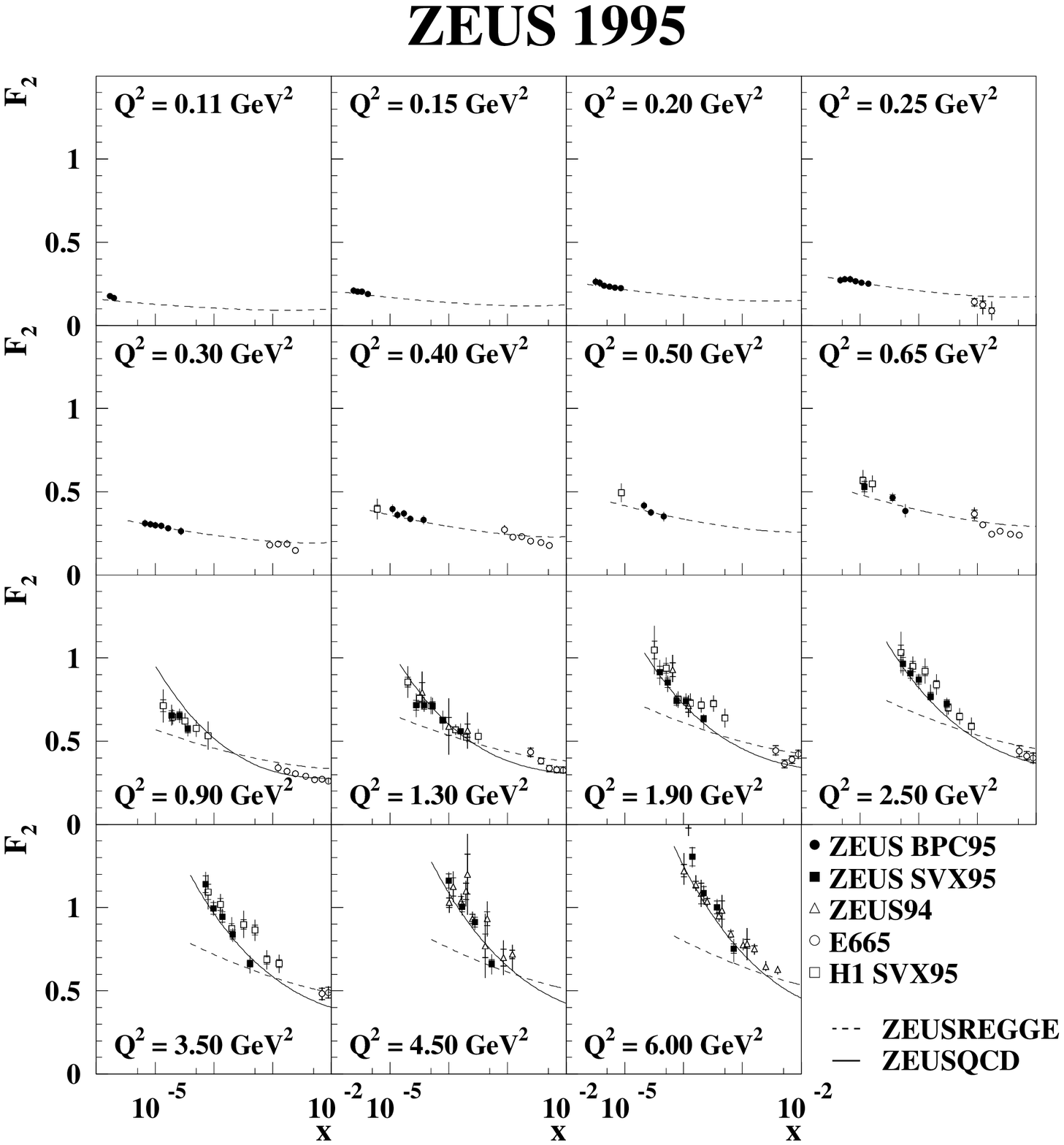,height=8cm}
\hfill}
\caption{Low $Q^2$ $F_2$ data for different $Q^2$ bins together with
the ZEUSDL style Regge model fit to the ZEUS BPC95 data. At larger
$Q^2$ values the ZEUS NLO QCD fit is also shown.}
\label{fig:svx95}
\end{figure}

Fig.~\ref{fig:svx95} shows the results for $F_2$ as a function of $x$
in bins of $Q^2$ between 0.65 and $6\;\rm GeV^2$ (ZEUS SVX95) together 
with ZEUS $F_2$ measurements at very low $Q^2 = 0.11 - 0.65\;\rm
GeV^2$ (ZEUS BPC95)
and at larger $Q^2$ those from the ZEUS94. There is good agreement
between the different ZEUS data sets in the region of overlap. Also
shown are data from the shifted vertex measurements by H1 (H1 SVX95)
and fixed target data from E665. The steep increase of $F_2$ at low
$x$ observed in the higher $Q^2$ bins softens at the lower $Q^2$
values of this analysis. The curves shown will be discussed later in
the text.

\subsection{The low $Q^2$ region}
We first give an overview of the low $Q^2$ region, $Q^2 < 5\;\rm
GeV^2$, taking ZEUS SVX95, BPC95 and ZEUS94 $F_2$ data. Using
Eq.~\ref{eqn:sigma_tot} we calculate $\sigma_{tot}^{\gamma^*p}$ values
from the $F_2$ data. The DL model predicts that the cross-section
rises slowly with energy $\propto W^{2\lambda}$, $\lambda = \alpha_P -
1 \approx 0.08$ and this behaviour seems to be followed by the data at
very low $Q^2$. Above $Q^2 = 0.65\;\rm GeV^2$, the DL model predicts a
shallower rise of the cross-section than the data exhibit. For $Q^2$
values of around $1\;\rm GeV^2$ and above, the GRV94 curves describe
the qualitative behaviour of the data, namely the increasing rise of
$\sigma_{tot}^{\gamma^*p}$ with $W^2$, as $Q^2$ increases. This
suggests that the perturbative QCD calculations can account for a
significant fraction of the cross-section at the larger $Q^2$ values.

\begin{figure}[ht]
\center{ \hfill
\psfig{figure=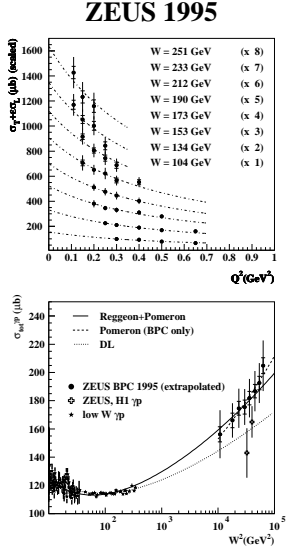,height=7.2cm} \hfill}
\caption{Upper plot: ZEUS BPC95 measurements of the total cross-section
$\sigma_T + \epsilon \sigma_L$ in bins of $W$ and the GVMD fit to the
data. Lower plot: $\sigma_{tot}^{\gamma p}$ as a function of
$W^2$. The ZEUS BPC95 points are those from the GVMD extrapolation of
$\sigma_0^{\gamma p}$.}
\label{fig:gvmd}
\end{figure}

For the remainder of this section we concentrate on non-perturbative
descriptions of the ZEUS BPC95 data.
Since BPC95 data are binned in $Q^2$ and $y$ we first
rewrite the double differential cross-section of
Eq.~\ref{eqn:doublediff} as $\frac{d^2 \sigma}{dy dQ^2} = \Gamma
\cdot (\sigma_T + \epsilon \sigma_L)$ where $\sigma_L = \frac{Q^2}{4 \pi^2
\alpha} F_L$ and $\sigma_T$ has been defined by
Eq.~\ref{eqn:sigma_tot}. The virtual photon has flux factor $\Gamma$
and polarisation $\epsilon$.
Keeping only the continuum states in the GVMD at a fixed $W$ the
longitudinal and transverse $\gamma^*p$ cross-section are related to
the corresponding photoproduction cross-section $\sigma_0^{\gamma p}$
by
\begin{eqnarray}
\sigma_L(W^2, Q^2) & = & \xi \left[ \frac{M_0^2}{Q^2}\ln\frac{M_0^2 +
Q^2}{M_0^2} - \frac{M_0^2}{M_0^2 + Q^2} \right] \sigma_0^{\gamma
p}(W^2) \nonumber\\
\sigma_T(W^2, Q^2) & = & \frac{M_0^2}{M_0^2 + Q^2} \sigma_0^{\gamma p} 
(W^2)
\end{eqnarray}
where the parameter $\xi$ is the ratio $\sigma_L^{Vp}/ \sigma_T^{Vp}$
for vector meson (V) proton scattering and $M_0$ is the effective
vector meson mass. Neither $\xi$ nor $M_0$ are given by the model and
they are either determined from a fit to data or by other
approaches. As we do not have much sensitivity to $\xi$ and it is
small (0.2 - 0.4) we set it here to zero. We thus have 9 parameters to
be determined by fitting the BPC data to the simplified GVMD
expression $F_2 =
\frac{Q^2 M_0^2}{M_0 + Q^2} \frac{\sigma_0^{\gamma p}}{4\pi^2 \alpha}$
in 8 bins of $W$ between 104 and 251 GeV. The fit is reasonable and
its quality might also be judged from the upper plot in
Fig.~\ref{fig:gvmd}. The value obtained for $M_0^2$ is $0.53 \pm 0.04
(stat) \pm 0.09 (sys)$. The resulting extrapolated values of
$\sigma_0^{\gamma p}$ are shown as a function of $W^2$ in the lower
plot of Fig.~\ref{fig:gvmd}, along with measurements from HERA and
lower energy experiments. The extrapolated BPC data lie somewhat above
the direct measurements from HERA. They are also above the cross
section prediction of the DL model.
It should be clearly
understood that the $\sigma_0^{\gamma p}$ data derived from the BPC95
data are not a measurement of the total photoproduction cross-section
but the result of a physically motivated ansatz.

The simple GVMD approach just described gives a concise account of the 
$Q^2$ dependence of the BPC data but it says nothing about the energy
dependence of $\sigma_0^{\gamma p}$. To explore this aspect of the
data we turn to a two component Regge model
\begin{eqnarray}
\sigma_{tot}^{\gamma p} (W^2) & = & A_R(W^2)^{\alpha_R-1} +
A_P(W^2)^{\alpha_P-1} \nonumber
\end{eqnarray}
where $P$ and $R$ denote the Pomeron and Reggeon contributions. The
Reggeon intercept $\alpha_R$ is fixed to the value 0.5 which is
compatible with the original DL value and by the re-evaluation of
Cudell et al. With such an intercept the Reggeon contribution is
negligible at HERA energies. Fitting the extrapolated BPC95 data alone
yields a value $1.141 \pm 0.020 (stat)$ for $\alpha_P$. Fitting both
terms to the real photoproduction data (with $W^2 > 3\;\rm GeV^2$) and 
BPC95 data yields $\alpha_P = 1.101 \pm 0.002 (stat)$. Including in
addition the two original measurements from HERA as well gives
$\alpha_P = 1.100 \pm 0.002 (stat)$. All these values of $\alpha_P$ are 
larger than the value of 1.08 used originally by DL, but we note that
the best estimate of Cudell et al. is 
$1.0964^{+0.0115}_{-0.0094}$,
which within the errors is consistent with our result.
The final step in the analysis of the BPC data is to combine the GVMD
fitted $Q^2$ dependence with the Regge model energy dependence
\begin{eqnarray}
\sigma_{tot}^{\gamma^* p} & = & \left( \frac{M_0^2}{M_0^2 +
Q^2}\right) (A_R (W^2)^{\alpha_R - 1} + A_P(W^2)^{\alpha_P -1}).\nonumber
\end{eqnarray} 
The parameters $M_0^2$ and $\alpha_R$ are fixed to their previous
values of 0.53 and 0.5, respectively. The 3 remaining parameters are
determined by fitting to real photoproduction data and the original
BPC data. The description of the low $Q^2$ $F_2$ data given by this DL
style model is shown in Fig.~\ref{fig:svx95}. Data in the BPC region
$Q^2 < 0.65\;\rm GeV^2$ is well described. At larger $Q^2$ values the
curves fall below the data. Also shown in Fig.~\ref{fig:svx95} for
$Q^2 > 6\;\rm GeV^2$ are the results of a NLO QCD fit (full line) as
described in Sec.~\ref{sec:qcdfit}.

\subsection{$F_2$ slopes: $d \ln F_2 / d \ln (1/x); dF_2 / d \ln Q^2$
\label{sec:slopes}}
To quantify the behaviour of $F_2$ as a function of $Q^2$ and $x$ at
low $x$ we calculate the two slopes $d \ln F_2 / d \ln (1/x); dF_2 /
d \ln Q^2$ from the ZEUS SVX95, BPC95 and ZEUS94 data sets.

At a fixed value of $Q^2$ and at small $x$ the behaviour of $F_2$ can
be characterised by $F_2 \propto x^{-\lambda}$, with $\lambda$ taking
rather different values in the Regge and BFKL
approaches. $\lambda_{eff}$ is calculated from horizontal slices of
ZEUS $F_2$ data between the $y = 1$ HERA kinematic limit and a fixed
cut of $x < 0.01$, here including E665 data. In a given $Q^2$ bin
$\langle\, x \,\rangle$ is calculated from the mean value of $\ln (1/x)$
weighted by the statistical errors of the corresponding $F_2$
values. The same procedure is applied to the theoretical curves shown
for comparison.

\begin{figure}[ht]
\vspace*{-1mm}
\center{
\hfill 
        \psfig{figure=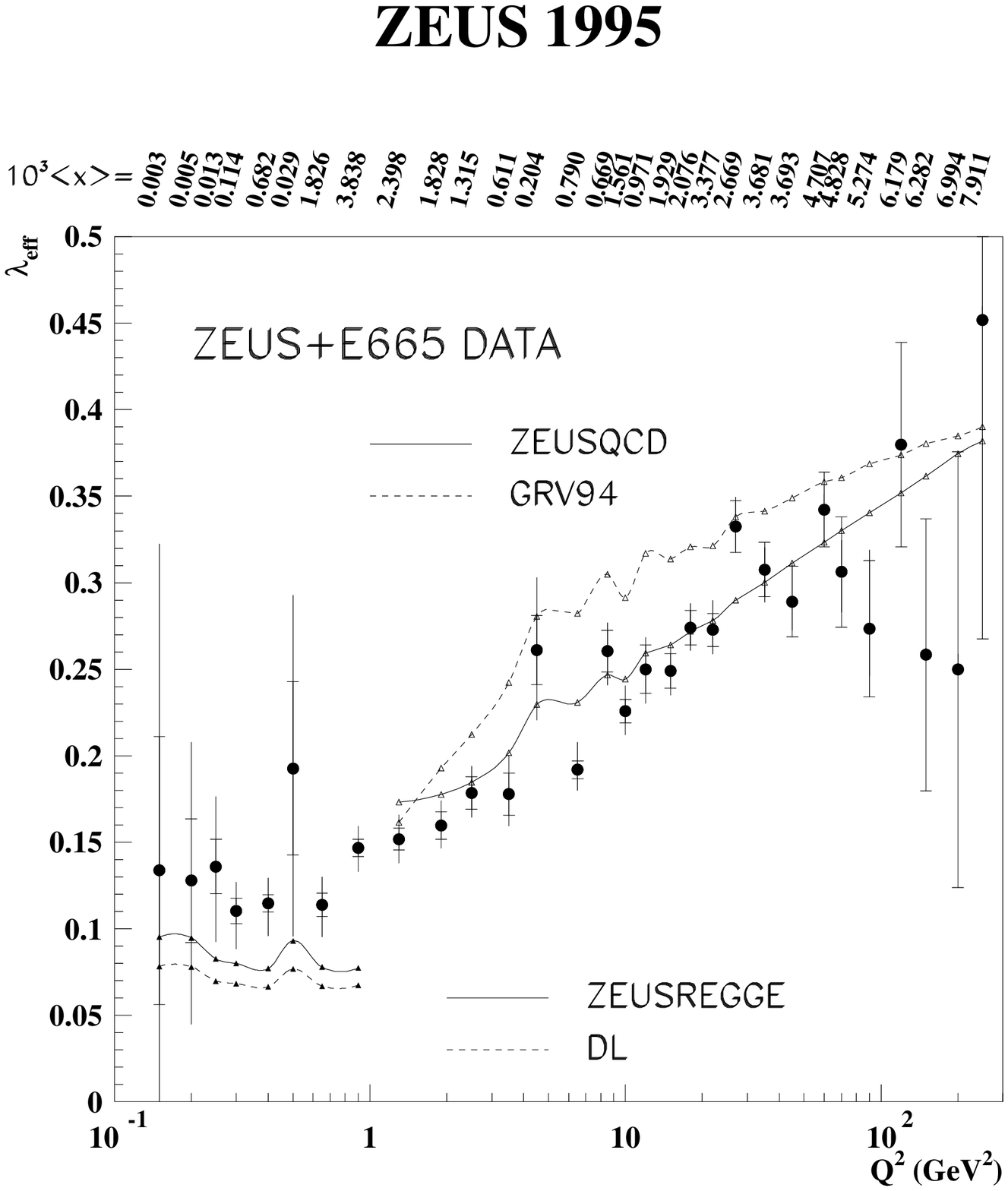,height=5cm}
        \hfill 
        \psfig{figure=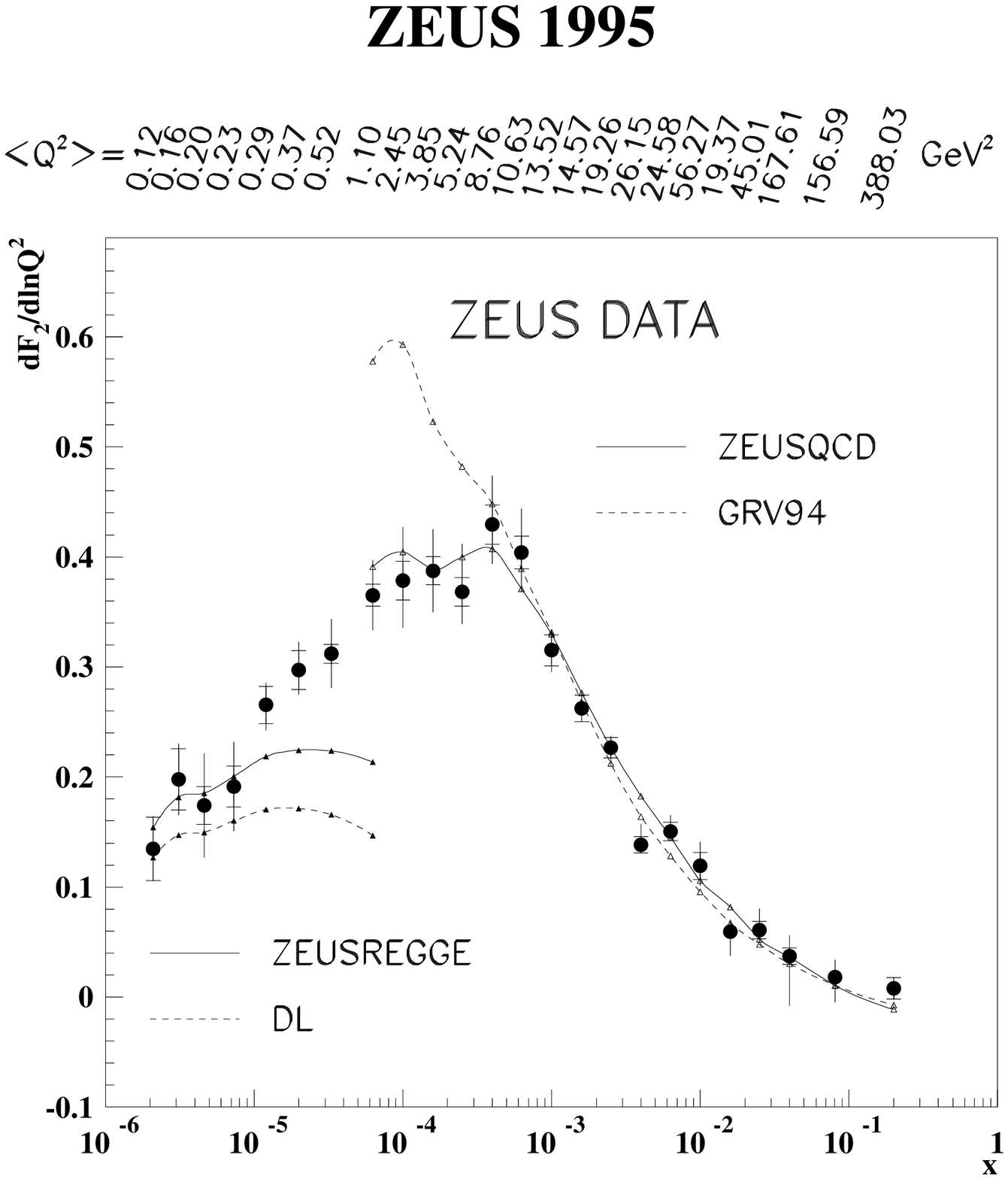,height=5cm}
\hfill }
\vspace*{-2mm}
\caption{Left plot: $d \ln F_2 / d \ln (1/x)$ as a function of $Q^2$
calculated by fitting ZEUS and E665 $F_2$ data in bins of $Q^2$. Right 
plot: $d F_2/ d \ln Q^2$ as a function of $x$ calculated by fitting
ZEUS $F_2$ data in bins of $x$.}
\label{fig:xq2_slopes}
\end{figure}

Figure \ref{fig:xq2_slopes} shows the measured values of $\lambda_{eff}$ as a 
function of $Q^2$. From the Regge approach one would expect
$\lambda_{eff} \approx 0.1$ and independent of $Q^2$. Data for $Q^2 <
1\;\rm GeV^2$ is consistent with this expectation. The linked points
labelled DL are calculated from the Donnachie-Landshoff fit
and as expected from the discussion of the previous section are
somewhat below the data. For $Q^2 > 1\;\rm GeV^2$, $\lambda_{eff}$
increases slowly to around 0.3 at $Q^2$ values of $40\;\rm
GeV^2$. Qualitatively the tendency of $\lambda_{eff}$ to increase with 
$Q^2$ is described by a number of pQCD approaches. The linked points
labelled GRV94 are calculated from the NLO QCD GRV94 fit. Although the 
GRV94 prediction follows the trend of the data it tends to lie above
the data, particularly in the $Q^2$ range $3 - 20\;\rm GeV^2$.

Within the framework of pQCD, at small $x$ the behaviour of $F_2$ is
largely determined by the behaviour of the sea quarks $F_2 \sim xS$,
whereas the $d F_2/d\ln Q^2$ is determined by the convolution of the
splitting function $P_{qg}$ and the gluon density, $dF_2/d\ln Q^2
\propto
\alpha_s P_{qg} \otimes g$. In order to study the scaling violations 
of $F_2$ in more detail the logarithmic slope $d F_2/d\ln Q^2$ is
derived from the data by fitting $F_2 = a + b\ln Q^2$ in bins of fixed
$x$. The statistical and systematic errors are determined as described
above. The results for $d F_2/d\ln Q^2$ as a function of $x$ are shown
in Fig.~\ref{fig:xq2_slopes}. For values of $x$ down to $3\times
10^{-4}$, the slopes are increasing as $x$ decreases. At lower values
of $x$ and $Q^2$, the slope decreases. Comparing the rapid increase in
$F_2$ at small $x$ with the behaviour of the $d F_2/d \ln Q^2$, one is
tempted to the naive conclusion that the underlying behaviour of the
sea quark and gluon momentum distributions must be different at small
$x$, with the sea dominant and the gluon tending to zero. The failure
of DL is in line with the earlier discussion. GRV94 does not follow
the trend of the data when it turns over.

\subsection{NLO QCD fit to $F_2$ data\label{sec:qcdfit}}
In perturbative QCD the scaling violations of the $F_2$ structure
function are caused by gluon bremsstrahlung from quarks and quark pair 
creation from gluons. In the low $x$ domain accessible at HERA the
latter process dominates the scaling violations. A QCD analysis of
$F_2$ structure functions measured at HERA therefore allows one to
extract the gluon momentum density in the proton down to low values of 
$x$. In this section we present NLO QCD fits to the ZEUS 1994 nominal
vertex data and the SVX95 data of this paper. We are not attempting to 
include all available information on parton densities, but
concentrating on what ZEUS data and their errors allow us to conclude
about the gluon density at low $x$.

To constrain the fits at high $x$ proton and deuteron $F_2$ structure
function data from NMC and BCDMS are included. The kinematic range
covered in this analysis is $3 \times 10^{-5} < x < 0.7$ and $1 < Q^2
< 5000\;\rm GeV^2$.

The QCD predictions for the $F_2$ structure functions are obtained by
solving the DGLAP evolution equations at NLO. These equations yield
the quark and gluon momentum distributions at all values of $Q^2$
provided they are given at some input scale $Q_0^2$. In this analysis
we adopt the so-called fixed flavour number scheme where only three
light flavours $(u, d, s)$ contribute to the quark density in the
proton. The corresponding structure functions $F_2^c$ and $F_2^b$ are
calculated from the photon-gluon fusion process including massive NLO
corrections. The input valence distributions are taken from the parton 
distribution set MRS(R2). As for MRS(R2) we assume that the strange
quark distribution is a given fraction $K_s = 0.2$ of the sea at the
scale $Q^2 = 1\;\rm GeV^2$. The gluon normalisation is fixed by the
momentum sum rule. The input value for the strong coupling constant is 
set to $\alpha_s(M_Z^2) = 0.118$ and the charm mass is taken to be
$m_c = 1.5\;\rm GeV$. In the QCD evolutions and the evaluation of the
structure functions the renormalisation scale and mass factorisation
scale are both set equal to $Q^2$. In the definition of the $\chi^2$
only statistical error are included and the relative normalisation of 
the data sets is fixed at unity. The fit yields a good description of
the data as shown in Fig.~\ref{fig:svx95}. We have also checked that
the gluon obtained from this fit to scaling violations is in agreement 
with the recent ZEUS measurements of charm production and $F_2^c$ in
deep inelastic scattering at HERA.

Two types of systematic uncertainties have been considered in this
analysis. `HERA standard errors' contain statistical error on the
data, experimental systematic uncertainties, relative normalisation
of the different data sets and uncertainties on $\alpha_s$, the
strange quark content of the proton and the charm
mass. `Parametrisation errors' contain uncertainties from a $\chi^2$
definition including statistical and experimental systematic errors,
variations of the starting scale $Q_0^2$ and an alternative, more
flexible parametrisation of the gluon density using Chebycheff
polynomials. The first type of errors amounts to $16\%$ $\Delta g/g$ 
at $x = 5 \times 10^{-5}$, $Q^2 = 7\;\rm GeV^2$, the second type
yields $9.5\%$ in $\Delta g/g$.

\begin{figure}[ht]
\vspace*{-3mm}
\center{
\hfill 
        \psfig{figure=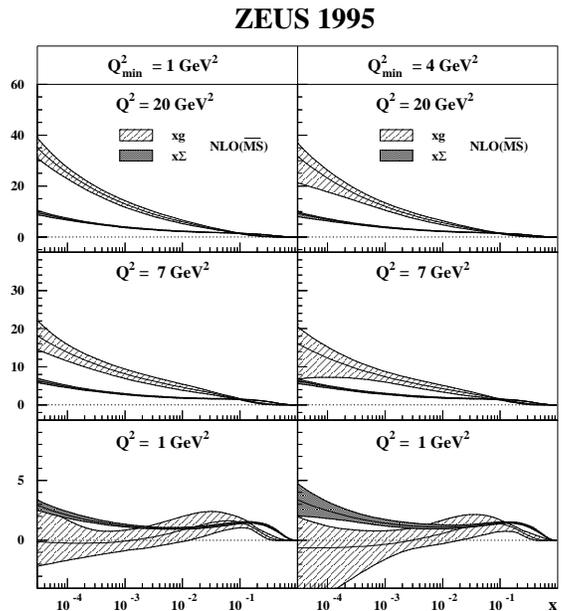,height=8cm}
\hfill }
\caption{The quark singlet momentum distribution, $x\Sigma$ (shaded),
and the gluon momentum distribution, $xg(x)$ (hatched), as a function
of $x$ at fixed values of $Q^2 = 1$, 7 and $20\;\rm GeV^2$. The error
bands correspond to the quadratic sum of all error sources considered
for each parton density.}
\label{fig:nlo_fit}
\end{figure}
The three plots of Fig.~\ref{fig:nlo_fit} show the distribution for
$x\Sigma$ and $xg$ as a function of $x$ for $Q^2$ at 1, 7 and $20\;\rm
GeV^2$. It can be seen that even at the smallest $Q^2$ $x\Sigma$ is
rising at small $x$ whereas the gluon distribution has become almost
flat. These results give support to the naive conclusion of
Sec.~\ref{sec:slopes}, that the sea distribution dominates at low $x$
and $Q^2$. At $Q^2 = 1\;\rm GeV^2$ the gluon distribution is poorly
determined and can, within errors, be negative at low $x$.


\section{Measurement of the Proton Structure Function $F_2$ from 1996
and 1997 data}

\subsection{Kinematics in Deep Inelastic Scattering}
Recalling the double differential NC
cross-section (\ref{eqn:doublediff}), but now including the corrections
($\delta_L \mbox{ and } \delta_3)$ for $F_L$ and $xF_3$ yields
\begin{eqnarray}
\frac{d^2 \sigma}{dx\, dQ^2} & = & \frac{2 \pi \alpha^2 Y_+}{x Q^4} F_2
\left( 1 - \delta_L - \delta_3 \right)
\left( 1 + \delta_r \right)
\end{eqnarray}
Here the $F_2$ structure function contains contributions from virtual
photon and $Z^0$ exchange
\begin{eqnarray}
F_2 & = & F_2^{em} + \frac{Q^2}{\left( Q^2 + M_Z^2 \right) } F_2^{int} +
\frac{Q^4}{\left( Q^2 + M_Z^2 \right)^2} F_2^{wk} 
 \end{eqnarray}
where $M_Z$ is the mass of the $Z^0$ and $F_2^{em}$, $F_2^{wk}$ and
$F_2^{int}$ are the contributions to $F_2$ due to photon exchange,
$Z^0$ exchange and $\gamma Z^0$ interference respectively. In this
analysis we determined the structure function $F_2^{em}$ using 1996
and 1997 data with an integrated luminosity of $6.8\;\rm pb^{-1}$ and
$27.4\;\rm pb^{-1}$, respectively.

The selection and kinematic reconstruction of NC DIS events is based
on an observed positron and the hadronic final state. For further
details see ref. \cite{vanc_f2}.

\subsection{Results}
Monte Carlo samples are used to estimate the acceptance, migration,
radiative corrections, electroweak corrections and background
contributions. $F_2^{em}$ is then determined based on a bin-by-bin
unfolding. The resulting statistical error, including the Monte Carlo
statistics, ranges from 2\% below $Q^2 = 100\;\rm GeV^2$ to 5-6\% at
$Q^2 \approx 800\;\rm GeV^2$.

The systematic uncertainties have been estimated by varying the
selection cuts, efficiencies and reconstruction techniques and
redetermining the cross section including background
estimates. Potential error source such as possible detector
misalignment, event vertex reconstruction, calorimeter energy scale,
positron identification efficiency, background contributions and
hadronic energy flow have been considered. The total systematic
uncertainty amounts to 3-4\% except at low and high $y$, where it
grows to 12\%.  At the present preliminary state of the analysis we
estimate an overall normalisation uncertainty of 3\%.

\begin{figure}[ht]
\center{\hbox{
        \psfig{figure=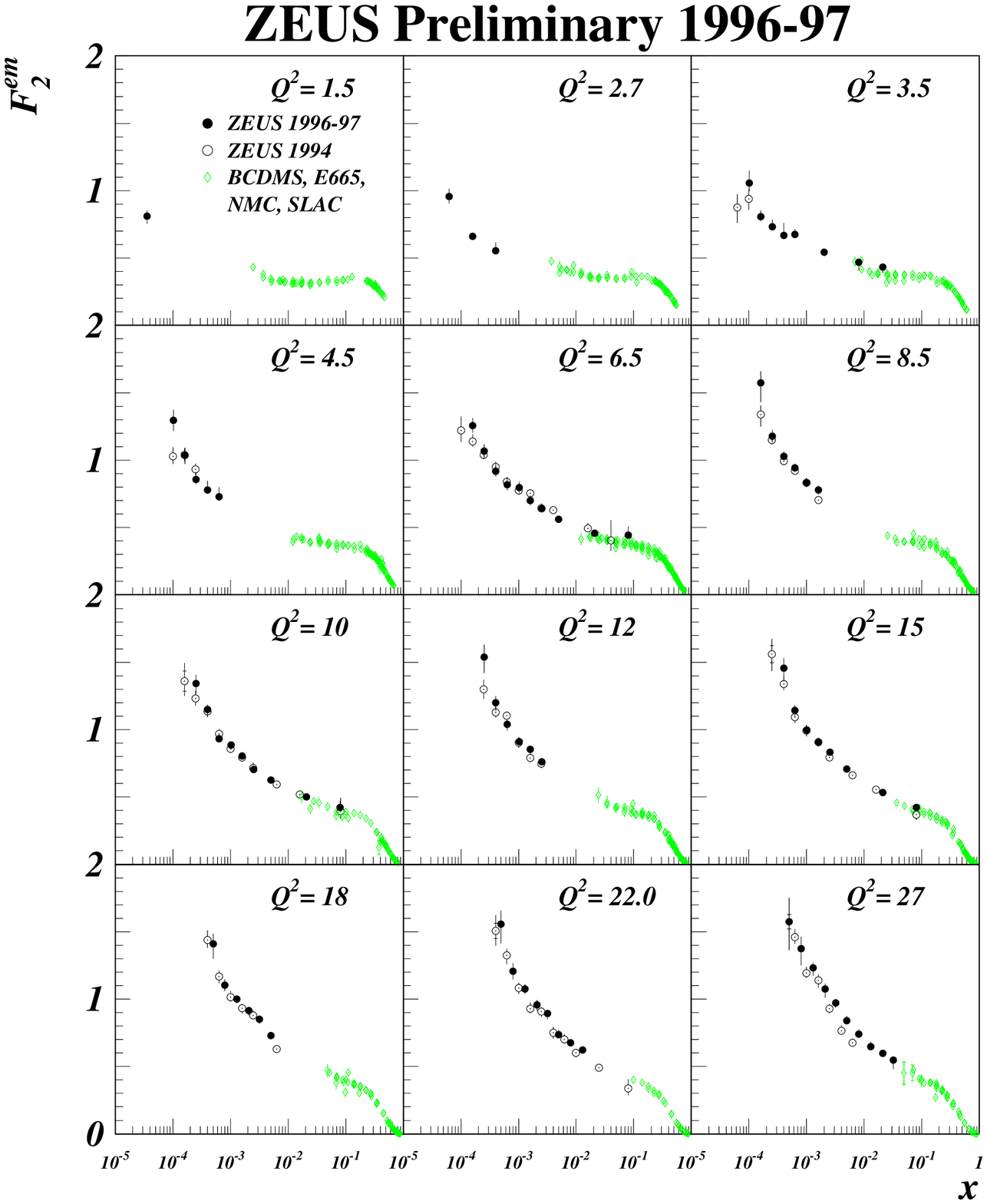,height=5.6cm}
        \psfig{figure=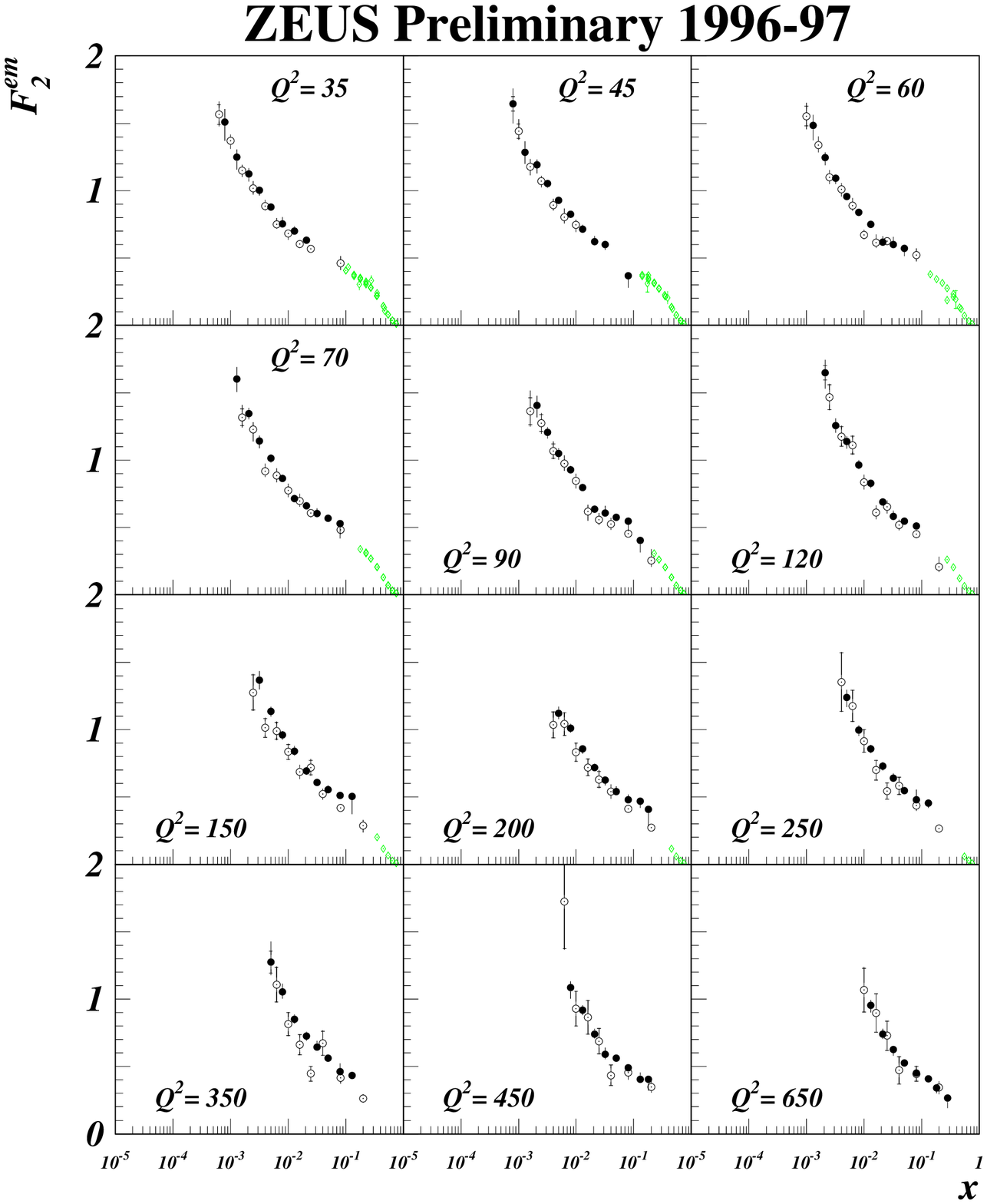,height=5.6cm}}}
        \vskip 0.2cm

\center{\hbox{
        \psfig{figure=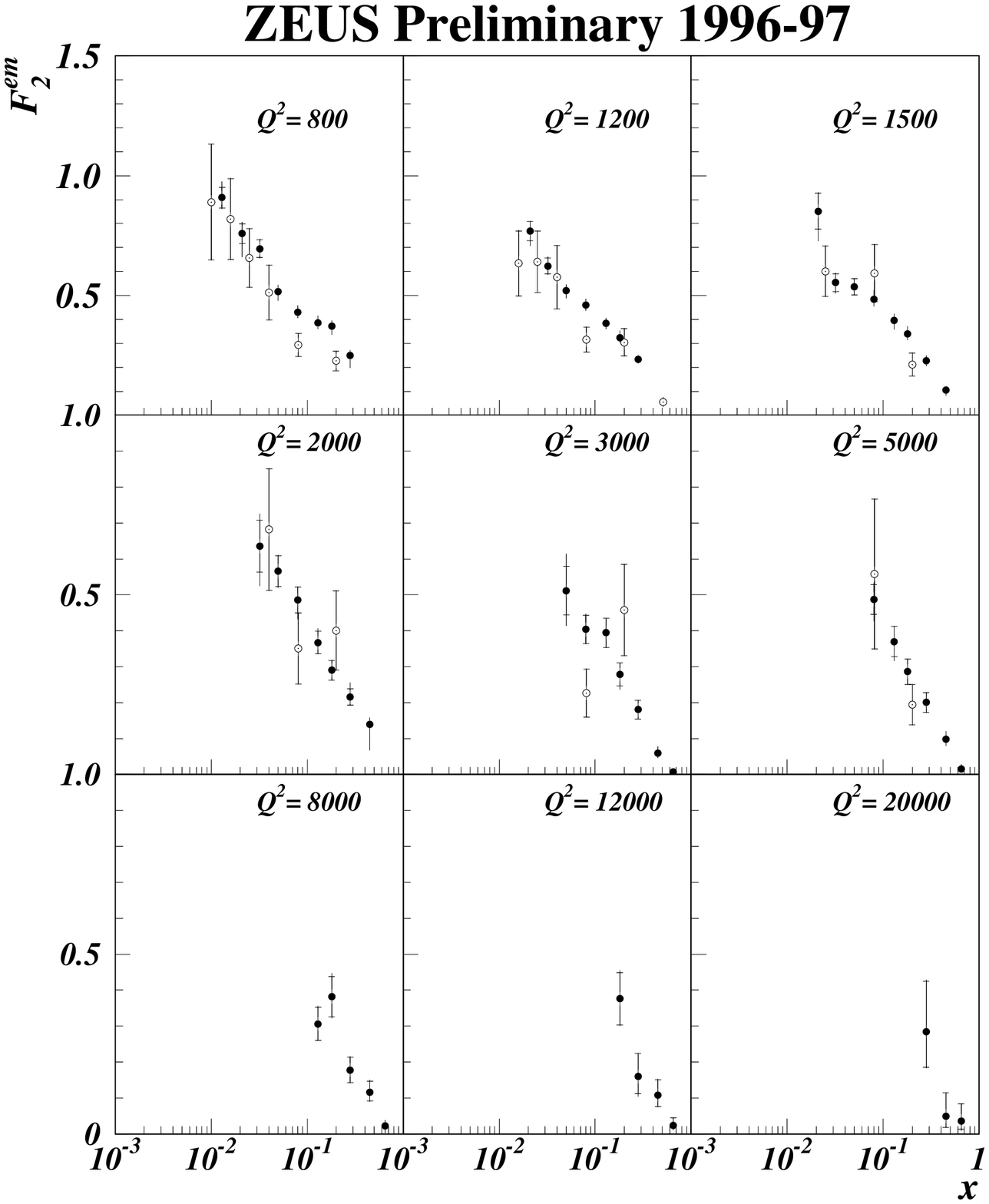,height=5.6cm}
        \hspace*{8mm}
        \psfig{figure=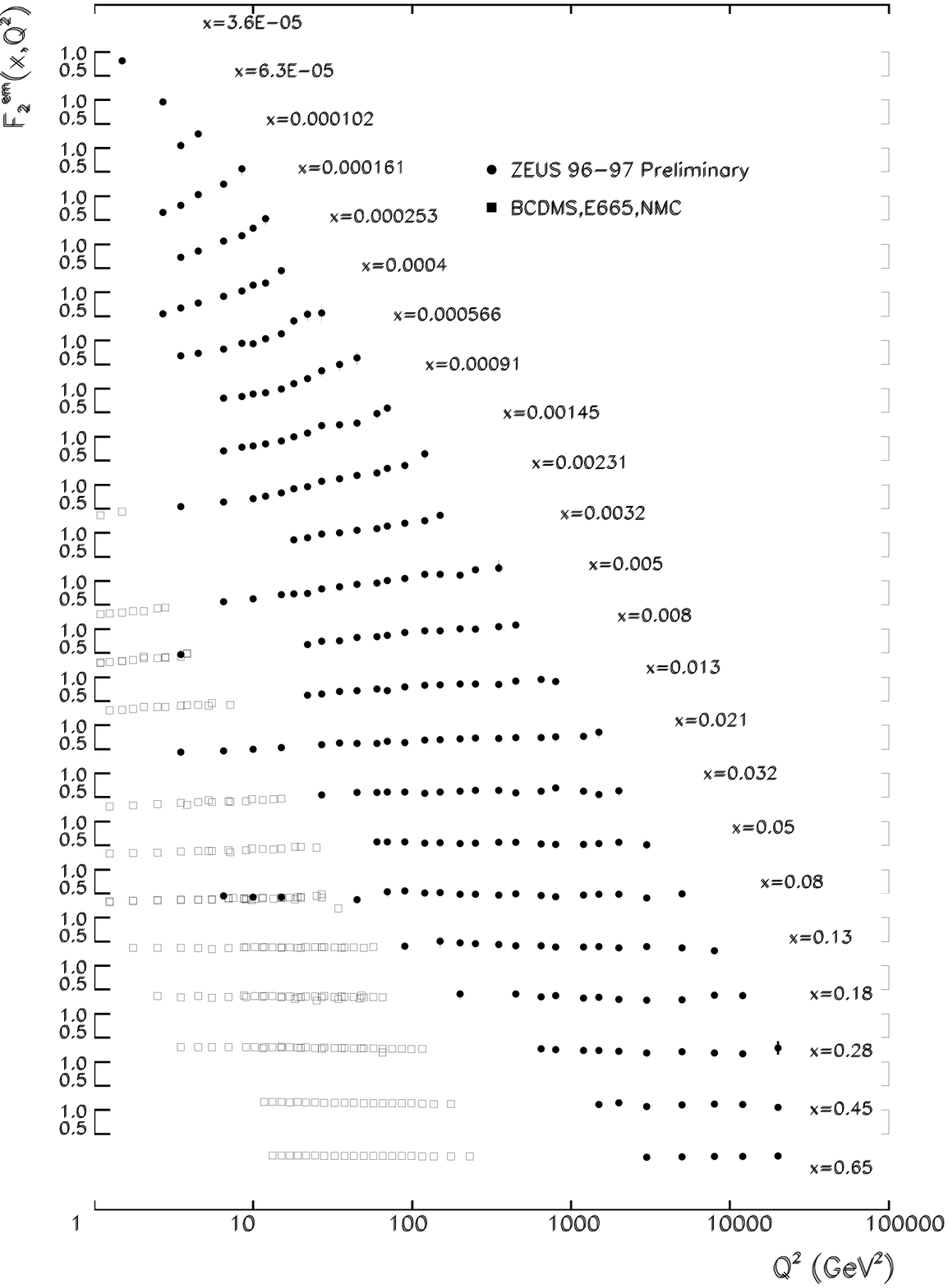,height=5.7cm}}}
\caption{Top and bottom left plots: $F_2^{em}$ versus $x$ for fixed
$Q^2$. Bottom right plot: $F_2^{em}$ as a function of $Q^2$ for fixed
$x$.}
\label{fig:f2x_1}
\end{figure}

The resulting $F_2^{em}$ is shown as a function of $x$ for fixed $Q^2$
in Figure~\ref{fig:f2x_1}.  Results from our previous analysis, and
from fixed target experiments are also shown for comparison. At low
$Q^2$ the rise in $F_2$ for $x
\rightarrow 0$ is measured with improved precision. The coverage in
$x$ has also been extended to higher $x$, yielding extended overlap
with the fixed target experiments; in the overlap region reasonable
agreement has been found. The $F_2$ scaling violation from this
analysis and the fixed target data are also shown in
Figure~\ref{fig:f2x_1}. For $Q^2 > 100\;\rm GeV^2$ the increase in
statistics allows a measurement of $F_2^{em}$ in smaller bins with
respect to our previous measurement. Above $Q^2 = 800\;\rm GeV^2$, the
statistical error grows typically to 5-15\% and dominates the total
error. Overall our data are in agreement with our published data.

\section*{References}
\begin{thebibliography}{99}

\bibitem{low_pheno}{\it ZEUS Results on the Measurement and
                           Phenomenology of $F_2$ at Low $x$ and Low
                           $Q^2$}, DESY 98-121 (August 1998) submitted
                           to The European Physical Journal.
\bibitem{vanc_f2}{\it Measurement of the Proton Structure Function
                         $F_2$ in $e^+p$ Collisions at HERA},
                         Submitted paper to the XXIX
                         International Conference on High Energy
                         Physics, Vancouver, July 23-29, 1998.
\end{thebibliography}
\end{document}